\documentclass[runningheads]{llncs}
\usepackage{latexsym}

\def\PsfigVersion{1.9}
\ifx\undefined\psfig\else \fi

\let\LaTeXAtSign=\@
\let\@=\relax
\edef\psfigRestoreAt{\catcode`\@=\number\catcode`@\relax}
\catcode`\@=11\relax
\newwrite\@unused
\def\ps@typeout#1{{\let\protect\string\immediate\write\@unused{#1}}}
\ps@typeout{psfig/tex \PsfigVersion}

\def\figurepath{./}

\def\@nnil{\@nil}
\def\@empty{}
\def\@psdonoop#1\@@#2#3{}
\def\@psdo#1:=#2\do#3{\edef\@psdotmp{#2}\ifx\@psdotmp\@empty \else
    \expandafter\@psdoloop#2,\@nil,\@nil\@@#1{#3}\fi}
\def\@psdoloop#1,#2,#3\@@#4#5{\def#4{#1}\ifx #4\@nnil \else
       #5\def#4{#2}\ifx #4\@nnil \else#5\@ipsdoloop #3\@@#4{#5}\fi\fi}
\def\@ipsdoloop#1,#2\@@#3#4{\def#3{#1}\ifx #3\@nnil 
       \let\@nextwhile=\@psdonoop \else
      #4\relax\let\@nextwhile=\@ipsdoloop\fi\@nextwhile#2\@@#3{#4}}
\def\@tpsdo#1:=#2\do#3{\xdef\@psdotmp{#2}\ifx\@psdotmp\@empty \else
    \@tpsdoloop#2\@nil\@nil\@@#1{#3}\fi}
\def\@tpsdoloop#1#2\@@#3#4{\def#3{#1}\ifx #3\@nnil 
       \let\@nextwhile=\@psdonoop \else
      #4\relax\let\@nextwhile=\@tpsdoloop\fi\@nextwhile#2\@@#3{#4}}
\ifx\undefined\fbox
\newdimen\fboxrule
\newdimen\fboxsep
\newdimen\ps@tempdima
\newbox\ps@tempboxa
\long\def\fbox#1{\leavevmode\setbox\ps@tempboxa\hbox{#1}\ps@tempdima\fboxrule
    \advance\ps@tempdima \fboxsep \advance\ps@tempdima \dp\ps@tempboxa
   \hbox{\lower \ps@tempdima\hbox
  {\vbox{\hrule height \fboxrule
          \hbox{\vrule width \fboxrule \hskip\fboxsep
          \vbox{\vskip\fboxsep \box\ps@tempboxa\vskip\fboxsep}\hskip 
                 \fboxsep\vrule width \fboxrule}
                 \hrule height \fboxrule}}}}
\fi
\newread\ps@stream
\newif\ifnot@eof       
\newif\if@noisy        
\newif\if@atend        
\newif\if@psfile       
{\catcode`\%=12\global\gdef\epsf@start{
\def\epsf@PS{PS}
\def\epsf@getbb#1{%
\openin\ps@stream=#1
\ifeof\ps@stream\ps@typeout{Error, File #1 not found}\else
   {\not@eoftrue \chardef\other=12
    \def\do##1{\catcode`##1=\other}\dospecials \catcode`\ =10
    \loop
       \if@psfile
	  \read\ps@stream to \epsf@fileline
       \else{
	  \obeyspaces
          \read\ps@stream to \epsf@tmp\global\let\epsf@fileline\epsf@tmp}
       \fi
       \ifeof\ps@stream\not@eoffalse\else
       \if@psfile\else
       \expandafter\epsf@test\epsf@fileline:. \\%
       \fi
          \expandafter\epsf@aux\epsf@fileline:. \\%
       \fi
   \ifnot@eof\repeat
   }\closein\ps@stream\fi}%
\long\def\epsf@test#1#2#3:#4\\{\def\epsf@testit{#1#2}
			\ifx\epsf@testit\epsf@start\else
\ps@typeout{Warning! File does not start with `\epsf@start'.  It may not be a PostScript file.}
			\fi
			\@psfiletrue} 
{\catcode`\%=12\global\let\epsf@percent=
\long\def\epsf@aux#1#2:#3\\{\ifx#1\epsf@percent
   \def\epsf@testit{#2}\ifx\epsf@testit\epsf@bblit
	\@atendfalse
        \epsf@atend #3 . \\%
	\if@atend	
	   \if@verbose{
		\ps@typeout{psfig: found `(atend)'; continuing search}
	   }\fi
        \else
        \epsf@grab #3 . . . \\%
        \not@eoffalse
        \global\no@bbfalse
        \fi
   \fi\fi}%
\def\epsf@grab #1 #2 #3 #4 #5\\{%
   \global\def\epsf@llx{#1}\ifx\epsf@llx\empty
      \epsf@grab #2 #3 #4 #5 .\\\else
   \global\def\epsf@lly{#2}%
   \global\def\epsf@urx{#3}\global\def\epsf@ury{#4}\fi}%
\def\epsf@atendlit{(atend)} 
\def\epsf@atend #1 #2 #3\\{%
   \def\epsf@tmp{#1}\ifx\epsf@tmp\empty
      \epsf@atend #2 #3 .\\\else
   \ifx\epsf@tmp\epsf@atendlit\@atendtrue\fi\fi}

\chardef\psletter = 11 
\chardef\other = 12

\newif \ifdebug 
\newif\ifc@mpute 
\c@mputetrue 

\let\then = \relax
\def\r@dian{pt }
\let\r@dians = \r@dian
\let\dimensionless@nit = \r@dian
\let\dimensionless@nits = \dimensionless@nit
\def\internal@nit{sp }
\let\internal@nits = \internal@nit
\newif\ifstillc@nverging
\def \Mess@ge #1{\ifdebug \then \message {#1} \fi}

{ 
	\catcode `\@ = \psletter
	\gdef \nodimen {\expandafter \n@dimen \the \dimen}
	\gdef \term #1 #2 #3%
	       {\edef \t@ {\the #1}
		\edef \t@@ {\expandafter \n@dimen \the #2\r@dian}%
		\t@rm {\t@} {\t@@} {#3}%
	       }
	\gdef \t@rm #1 #2 #3%
	       {{%
		\count 0 = 0
		\dimen 0 = 1 \dimensionless@nit
		\dimen 2 = #2\relax
		\Mess@ge {Calculating term #1 of \nodimen 2}%
		\loop
		\ifnum	\count 0 < #1
		\then	\advance \count 0 by 1
			\Mess@ge {Iteration \the \count 0 \space}%
			\Multiply \dimen 0 by {\dimen 2}%
			\Mess@ge {After multiplication, term = \nodimen 0}%
			\Divide \dimen 0 by {\count 0}%
			\Mess@ge {After division, term = \nodimen 0}%
		\repeat
		\Mess@ge {Final value for term #1 of 
				\nodimen 2 \space is \nodimen 0}%
		\xdef \Term {#3 = \nodimen 0 \r@dians}%
		\aftergroup \Term
	       }}
	\catcode `\p = \other
	\catcode `\t = \other
	\gdef \n@dimen #1pt{#1} 
}

\def \Divide #1by #2{\divide #1 by #2} 

\def \Multiply #1by #2
       {{
	\count 0 = #1\relax
	\count 2 = #2\relax
	\count 4 = 65536
	\Mess@ge {Before scaling, count 0 = \the \count 0 \space and
			count 2 = \the \count 2}%
	\ifnum	\count 0 > 32767 
	\then	\divide \count 0 by 4
		\divide \count 4 by 4
	\else	\ifnum	\count 0 < -32767
		\then	\divide \count 0 by 4
			\divide \count 4 by 4
		\else
		\fi
	\fi
	\ifnum	\count 2 > 32767 
	\then	\divide \count 2 by 4
		\divide \count 4 by 4
	\else	\ifnum	\count 2 < -32767
		\then	\divide \count 2 by 4
			\divide \count 4 by 4
		\else
		\fi
	\fi
	\multiply \count 0 by \count 2
	\divide \count 0 by \count 4
	\xdef \product {#1 = \the \count 0 \internal@nits}%
	\aftergroup \product
       }}

\def\r@duce{\ifdim\dimen0 > 90\r@dian \then   
		\multiply\dimen0 by -1
		\advance\dimen0 by 180\r@dian
		\r@duce
	    \else \ifdim\dimen0 < -90\r@dian \then  
		\advance\dimen0 by 360\r@dian
		\r@duce
		\fi
	    \fi}

\def\Sine#1%
       {{%
	\dimen 0 = #1 \r@dian
	\r@duce
	\ifdim\dimen0 = -90\r@dian \then
	   \dimen4 = -1\r@dian
	   \c@mputefalse
	\fi
	\ifdim\dimen0 = 90\r@dian \then
	   \dimen4 = 1\r@dian
	   \c@mputefalse
	\fi
	\ifdim\dimen0 = 0\r@dian \then
	   \dimen4 = 0\r@dian
	   \c@mputefalse
	\fi
	\ifc@mpute \then
		\divide\dimen0 by 180
		\dimen0=3.141592654\dimen0
		\dimen 2 = 3.1415926535897963\r@dian 
		\divide\dimen 2 by 2 
		\Mess@ge {Sin: calculating Sin of \nodimen 0}%
		\count 0 = 1 
		\dimen 2 = 1 \r@dian 
		\dimen 4 = 0 \r@dian 
		\loop
			\ifnum	\dimen 2 = 0 
			\then	\stillc@nvergingfalse 
			\else	\stillc@nvergingtrue
			\fi
			\ifstillc@nverging 
			\then	\term {\count 0} {\dimen 0} {\dimen 2}%
				\advance \count 0 by 2
				\count 2 = \count 0
				\divide \count 2 by 2
				\ifodd	\count 2 
				\then	\advance \dimen 4 by \dimen 2
				\else	\advance \dimen 4 by -\dimen 2
				\fi
		\repeat
	\fi		
			\xdef \sine {\nodimen 4}%
       }}

\def\Cosine#1{\ifx\sine\UnDefined\edef\Savesine{\relax}\else
		             \edef\Savesine{\sine}\fi
	{\dimen0=#1\r@dian\advance\dimen0 by 90\r@dian
	 \Sine{\nodimen 0}
	 \xdef\cosine{\sine}
	 \xdef\sine{\Savesine}}}	      

\def\psdraft{
	\def\@psdraft{0}
}
\def\psfull{
	\def\@psdraft{100}
}

\psfull

\newif\if@scalefirst
\def\psscalefirst{\@scalefirsttrue}
\def\psrotatefirst{\@scalefirstfalse}
\psrotatefirst

\newif\if@draftbox
\def\psnodraftbox{
	\@draftboxfalse
}
\def\psdraftbox{
	\@draftboxtrue
}
\@draftboxtrue

\newif\if@prologfile
\newif\if@postlogfile
\def\pssilent{
	\@noisyfalse
}
\def\psnoisy{
	\@noisytrue
}
\psnoisy
\newif\if@bbllx
\newif\if@bblly
\newif\if@bburx
\newif\if@bbury
\newif\if@height
\newif\if@width
\newif\if@rheight
\newif\if@rwidth
\newif\if@angle
\newif\if@clip
\newif\if@verbose
\def\@p@@sclip#1{\@cliptrue}

\newif\if@decmpr

\def\@p@@sfigure#1{\def\@p@sfile{null}\def\@p@sbbfile{null}
	        \openin1=#1.bb
		\ifeof1\closein1
	        	\openin1=\figurepath#1.bb
			\ifeof1\closein1
			        \openin1=#1
				\ifeof1\closein1%
				       \openin1=\figurepath#1
					\ifeof1
					   \ps@typeout{Error, File #1 not found}
						\if@bbllx\if@bblly
				   		\if@bburx\if@bbury
			      				\def\@p@sfile{#1}%
			      				\def\@p@sbbfile{#1}%
							\@decmprfalse
				  	   	\fi\fi\fi\fi
					\else\closein1
				    		\def\@p@sfile{\figurepath#1}%
				    		\def\@p@sbbfile{\figurepath#1}%
						\@decmprfalse
	                       		\fi%
			 	\else\closein1%
					\def\@p@sfile{#1}
					\def\@p@sbbfile{#1}
					\@decmprfalse
			 	\fi
			\else
				\def\@p@sfile{\figurepath#1}
				\def\@p@sbbfile{\figurepath#1.bb}
				\@decmprtrue
			\fi
		\else
			\def\@p@sfile{#1}
			\def\@p@sbbfile{#1.bb}
			\@decmprtrue
		\fi}

\def\@p@@sfile#1{\@p@@sfigure{#1}}

\def\@p@@sbbllx#1{
		\@bbllxtrue
		\dimen100=#1
		\edef\@p@sbbllx{\number\dimen100}
}
\def\@p@@sbblly#1{
		\@bbllytrue
		\dimen100=#1
		\edef\@p@sbblly{\number\dimen100}
}
\def\@p@@sbburx#1{
		\@bburxtrue
		\dimen100=#1
		\edef\@p@sbburx{\number\dimen100}
}
\def\@p@@sbbury#1{
		\@bburytrue
		\dimen100=#1
		\edef\@p@sbbury{\number\dimen100}
}
\def\@p@@sheight#1{
		\@heighttrue
		\dimen100=#1
   		\edef\@p@sheight{\number\dimen100}
}
\def\@p@@swidth#1{
		\@widthtrue
		\dimen100=#1
		\edef\@p@swidth{\number\dimen100}
}
\def\@p@@srheight#1{
		\@rheighttrue
		\dimen100=#1
		\edef\@p@srheight{\number\dimen100}
}
\def\@p@@srwidth#1{
		\@rwidthtrue
		\dimen100=#1
		\edef\@p@srwidth{\number\dimen100}
}
\def\@p@@sangle#1{
		\@angletrue
		\edef\@p@sangle{#1} 
}
\def\@p@@ssilent#1{ 
		\@verbosefalse
}
\def\@p@@sprolog#1{\@prologfiletrue\def\@prologfileval{#1}}
\def\@p@@spostlog#1{\@postlogfiletrue\def\@postlogfileval{#1}}
\def\@cs@name#1{\csname #1\endcsname}
\def\@setparms#1=#2,{\@cs@name{@p@@s#1}{#2}}
\def\ps@init@parms{
		\@bbllxfalse \@bbllyfalse
		\@bburxfalse \@bburyfalse
		\@heightfalse \@widthfalse
		\@rheightfalse \@rwidthfalse
		\def\@p@sbbllx{}\def\@p@sbblly{}
		\def\@p@sbburx{}\def\@p@sbbury{}
		\def\@p@sheight{}\def\@p@swidth{}
		\def\@p@srheight{}\def\@p@srwidth{}
		\def\@p@sangle{0}
		\def\@p@sfile{} \def\@p@sbbfile{}
		\def\@p@scost{10}
		\def\@sc{}
		\@prologfilefalse
		\@postlogfilefalse
		\@clipfalse
		\if@noisy
			\@verbosetrue
		\else
			\@verbosefalse
		\fi
}
\def\parse@ps@parms#1{
	 	\@psdo\@psfiga:=#1\do
		   {\expandafter\@setparms\@psfiga,}}
\newif\ifno@bb
\def\bb@missing{
	\if@verbose{
		\ps@typeout{psfig: searching \@p@sbbfile \space  for bounding box}
	}\fi
	\no@bbtrue
	\epsf@getbb{\@p@sbbfile}
        \ifno@bb \else \bb@cull\epsf@llx\epsf@lly\epsf@urx\epsf@ury\fi
}	
\def\bb@cull#1#2#3#4{
	\dimen100=#1 bp\edef\@p@sbbllx{\number\dimen100}
	\dimen100=#2 bp\edef\@p@sbblly{\number\dimen100}
	\dimen100=#3 bp\edef\@p@sbburx{\number\dimen100}
	\dimen100=#4 bp\edef\@p@sbbury{\number\dimen100}
	\no@bbfalse
}
\newdimen\p@intvaluex
\newdimen\p@intvaluey
\def\rotate@#1#2{{\dimen0=#1 sp\dimen1=#2 sp
		  \global\p@intvaluex=\cosine\dimen0
		  \dimen3=\sine\dimen1
		  \global\advance\p@intvaluex by -\dimen3
		  \global\p@intvaluey=\sine\dimen0
		  \dimen3=\cosine\dimen1
		  \global\advance\p@intvaluey by \dimen3
		  }}
\def\compute@bb{
		\no@bbfalse
		\if@bbllx \else \no@bbtrue \fi
		\if@bblly \else \no@bbtrue \fi
		\if@bburx \else \no@bbtrue \fi
		\if@bbury \else \no@bbtrue \fi
		\ifno@bb \bb@missing \fi
		\ifno@bb \ps@typeout{FATAL ERROR: no bb supplied or found}
			\no-bb-error
		\fi
		\count203=\@p@sbburx
		\count204=\@p@sbbury
		\advance\count203 by -\@p@sbbllx
		\advance\count204 by -\@p@sbblly
		\edef\ps@bbw{\number\count203}
		\edef\ps@bbh{\number\count204}
		\if@angle 
			\Sine{\@p@sangle}\Cosine{\@p@sangle}
	        	{\dimen100=\maxdimen\xdef\r@p@sbbllx{\number\dimen100}
					    \xdef\r@p@sbblly{\number\dimen100}
			                    \xdef\r@p@sbburx{-\number\dimen100}
					    \xdef\r@p@sbbury{-\number\dimen100}}
                        \def\minmaxtest{
			   \ifnum\number\p@intvaluex<\r@p@sbbllx
			      \xdef\r@p@sbbllx{\number\p@intvaluex}\fi
			   \ifnum\number\p@intvaluex>\r@p@sbburx
			      \xdef\r@p@sbburx{\number\p@intvaluex}\fi
			   \ifnum\number\p@intvaluey<\r@p@sbblly
			      \xdef\r@p@sbblly{\number\p@intvaluey}\fi
			   \ifnum\number\p@intvaluey>\r@p@sbbury
			      \xdef\r@p@sbbury{\number\p@intvaluey}\fi
			   }
			\rotate@{\@p@sbbllx}{\@p@sbblly}
			\minmaxtest
			\rotate@{\@p@sbbllx}{\@p@sbbury}
			\minmaxtest
			\rotate@{\@p@sbburx}{\@p@sbblly}
			\minmaxtest
			\rotate@{\@p@sbburx}{\@p@sbbury}
			\minmaxtest
			\edef\@p@sbbllx{\r@p@sbbllx}\edef\@p@sbblly{\r@p@sbblly}
			\edef\@p@sbburx{\r@p@sbburx}\edef\@p@sbbury{\r@p@sbbury}
		\fi
		\count203=\@p@sbburx
		\count204=\@p@sbbury
		\advance\count203 by -\@p@sbbllx
		\advance\count204 by -\@p@sbblly
		\edef\@bbw{\number\count203}
		\edef\@bbh{\number\count204}
}
\def\in@hundreds#1#2#3{\count240=#2 \count241=#3
		     \count100=\count240	
		     \divide\count100 by \count241
		     \count101=\count100
		     \multiply\count101 by \count241
		     \advance\count240 by -\count101
		     \multiply\count240 by 10
		     \count101=\count240	
		     \divide\count101 by \count241
		     \count102=\count101
		     \multiply\count102 by \count241
		     \advance\count240 by -\count102
		     \multiply\count240 by 10
		     \count102=\count240	
		     \divide\count102 by \count241
		     \count200=#1\count205=0
		     \count201=\count200
			\multiply\count201 by \count100
		 	\advance\count205 by \count201
		     \count201=\count200
			\divide\count201 by 10
			\multiply\count201 by \count101
			\advance\count205 by \count201
		     \count201=\count200
			\divide\count201 by 100
			\multiply\count201 by \count102
			\advance\count205 by \count201
		     \edef\@result{\number\count205}
}
\def\compute@wfromh{
		\in@hundreds{\@p@sheight}{\@bbw}{\@bbh}
		\edef\@p@swidth{\@result}
}
\def\compute@hfromw{
	        \in@hundreds{\@p@swidth}{\@bbh}{\@bbw}
		\edef\@p@sheight{\@result}
}
\def\compute@handw{
		\if@height 
			\if@width
			\else
				\compute@wfromh
			\fi
		\else 
			\if@width
				\compute@hfromw
			\else
				\edef\@p@sheight{\@bbh}
				\edef\@p@swidth{\@bbw}
			\fi
		\fi
}
\def\compute@resv{
		\if@rheight \else \edef\@p@srheight{\@p@sheight} \fi
		\if@rwidth \else \edef\@p@srwidth{\@p@swidth} \fi
}
\def\compute@sizes{
	\compute@bb
	\if@scalefirst\if@angle
	\if@width
	   \in@hundreds{\@p@swidth}{\@bbw}{\ps@bbw}
	   \edef\@p@swidth{\@result}
	\fi
	\if@height
	   \in@hundreds{\@p@sheight}{\@bbh}{\ps@bbh}
	   \edef\@p@sheight{\@result}
	\fi
	\fi\fi
	\compute@handw
	\compute@resv}

\def\psfig#1{\vbox {
	\ps@init@parms
	\parse@ps@parms{#1}
	\compute@sizes
	\ifnum\@p@scost<\@psdraft{
		\special{ps::[begin] 	\@p@swidth \space \@p@sheight \space
				\@p@sbbllx \space \@p@sbblly \space
				\@p@sbburx \space \@p@sbbury \space
				startTexFig \space }
		\if@angle
			\special {ps:: \@p@sangle \space rotate \space} 
		\fi
		\if@clip{
			\if@verbose{
				\ps@typeout{(clip)}
			}\fi
			\special{ps:: doclip \space }
		}\fi
		\if@prologfile
		    \special{ps: plotfile \@prologfileval \space } \fi
		\if@decmpr{
			\if@verbose{
				\ps@typeout{psfig: including \@p@sfile.Z \space }
			}\fi
			\special{ps: plotfile "`zcat \@p@sfile.Z" \space }
		}\else{
			\if@verbose{
				\ps@typeout{psfig: including \@p@sfile \space }
			}\fi
			\special{ps: plotfile \@p@sfile \space }
		}\fi
		\if@postlogfile
		    \special{ps: plotfile \@postlogfileval \space } \fi
		\special{ps::[end] endTexFig \space }
		\vbox to \@p@srheight sp{
			\hbox to \@p@srwidth sp{
				\hss
			}
		\vss
		}
	}\else{
		\if@draftbox{		
			\hbox{\frame{\vbox to \@p@srheight sp{
			\vss
			\hbox to \@p@srwidth sp{ \hss \@p@sfile \hss }
			\vss
			}}}
		}\else{
			\vbox to \@p@srheight sp{
			\vss
			\hbox to \@p@srwidth sp{\hss}
			\vss
			}
		}\fi

	}\fi
}}
\psfigRestoreAt
\let\@=\LaTeXAtSign

\def\punto{\hspace*{\fill}\Box}

\pagestyle{empty}
\date{}
\begin{document}

\title{X-Learn: an XML-based, multi-agent system for supporting ``user-device'' adaptive e-learning}

\titlerunning{X-Learn: an XML-based, multi-agent system ...}


\author{Pasquale De Meo\inst{1} \and Alfredo Garro\inst{2} \and \\
Giorgio Terracina\inst{3} \and Domenico Ursino\inst{1}}
\authorrunning{Alfredo Garro \and Pasquale De Meo \and Giorgio Terracina \and Domenico Ursino}
\tocauthor{Pasquale De Meo (DIMET, Universit\`a Mediterranea di Reggio Calabria), Alfredo Garro
(DEIS, Universit\`a della Calabria), Giorgio Terracina (Dip. di Matematica, Universit\`a della
Calabria), Domenico Ursino (DIMET, Universit\`a Mediterranea di Reggio Calabria)} \institute{
DIMET, Universit\`a Mediterranea di Reggio Calabria, Via Graziella, Localit\`a Feo di Vito, 89060
Reggio Calabria, Italy, \and DEIS, Universit\`a della Calabria, Via Pietro Bucci, 87036 Rende (CS),
Italy \and Dipartimento di Matematica, Universit\`a della Calabria, Via Pietro Bucci,
87036 Rende (CS), Italy\\
\email{demeo@ing.unirc.it, garro@deis.unical.it, \\ terracina@mat.unical.it, ursino@unirc.it}}

\maketitle

\begin{abstract}

In this paper we present {\em X-Learn}, an XML-based, multi-agent system for supporting
``user-device'' adaptive e-learning. {\em X-Learn} is characterized by the following features: {\em
(i)} it is highly subjective, since it handles quite a rich and detailed user profile that plays a
key role during the learning activities; {\em (ii)} it is dynamic and flexible, i.e., it is capable
of reacting to variations of exigencies and objectives; {\em (iii)} it is device-adaptive, since it
decides the learning objects to present to the user on the basis of the device she/he is currently
exploiting; {\em (iv)} it is generic, i.e., it is capable of operating in a large variety of
learning contexts; {\em (v)} it is XML based, since it exploits many facilities of XML technology
for handling and exchanging information connected to e-learning activities. The paper reports also
various experimental results as well as a comparison between {\em X-Learn} and other related
e-learning management systems already presented in the literature.

\end{abstract}

\section{Introduction}
\label{Introduction}

E-learning can be defined as the activity that supports a learning experience by either developing
or applying Information \& Communication Technology (ICT). It is playing a more and more relevant
role in the ICT market and its importance is becoming crucial for organizing training in
businesses. Indeed, market dynamism compels organizations to avoid medium-to-long term programming
and to work in a project-shaped, short-to-medium term perspective.

In order to realize projects which it is involved in, an organization continuously needs new ``know
how'' and competences; owing to the growing ``skill shortage'', these can be found on the external
market only with a great difficulty and a high cost. As a consequence, the capability to internally
construct the necessary know how has become a must for an organization.

E-learning is a particularly suitable solution to these exigencies. More specifically, an
e-learning platform should initially determine the competence gap of the human resources assigned
to a project; after this, it should fill such a gap by constructing suitable personalized and
flexible learning programs that can be dinamically adapted to the feedback received by the user.

In such a context, in order to guarantee the maximum flexibility and, contemporarily, the highest
efficiency to e-learning activities, it has been proposed to organize learning contents into
independent units, named {\em learning objects}, that can be dynamically combined for constructing
personalized learning programs. In order to successfully perform such an activity, an efficient and
effective organization of available learning objects appears crucial. In other words, it appears
necessary to define and construct a meta-knowledge that allows to classify available learning
objects (documents, slides, simulations, role games, questionnaires, tests, registered lessons,
etc.) on the basis of their objectives, arguments, exploited media and so on.

In order to both simplify learning object exploitation and foster platform interoperability,
important international organizations have proposed to associate suitable descriptors, named LOM
(Learning Object Metadata), with learning objects \cite{LOM}. LOM allow information about learning
objects to be obtained without the necessity to directly analyze them. More specifically, the
Instruction Management System (IMS) \cite{IMS}, an authoritative organism for LOM standardization,
has proposed to describe learning objects by means of an XML document which a suitable XML Schema
is associated with. Such a proposal has been favourably accepted by the e-learning community and,
presently, almost all commercial e-learning platforms support it.

LOM paradigm has largely facilitated e-learning activities, in particular the automatic
construction of learning programs. However, in order to improve the efficiency and the
effectiveness of e-learning activities, some important problems, often involving research areas
quite far from computer science, should be faced. As an example, new didactic methodologies, based
on the learning object paradigm and well suited to automatically realize learning programs, should
be defined \cite{Wiley01}. In addition, a continuous and pervasive e-learning activity should
carefully consider and support the different devices that users might exploit during their learning
process. With regard to this, it is worth observing that, in the Personal Digital Assistant era,
limiting users to perform e-learning activities only by a Personal Computer connected to the
organization's LAN unjustifiably reduces the flexibility and, consequently, both the efficiency and
the effectiveness of the e-learning process.

In our opinion, some of these challenging issues can be successfully faced by exploiting the agent
technology. The present paper aims at showing the feasibility of this idea; in particular, it
presents {\em X-Learn}, an XML-based Multi-Agent System for supporting ``user-device adaptive''
e-learning. More specifically, X-Learn has been conceived for assisting users to learn new ``know
how'' and competences to fill the gap between their present knowledge and that required by a new
project which they have been assigned to.

In {\em X-Learn} user assistance is guaranteed by constructing personalized, flexible and dynamic
learning programs taking into account the background knowledge of a user, her/his didactic
objectives as well as devices and connection typologies she/he intends to exploit for carrying out
her/his e-learning activity.

{\em X-Learn} is characterized by the following features that, in our opinion, are extremely
relevant for a new e-learning system:

\begin{itemize}

\item {\em It is highly subjective}; indeed, it handles quite a rich and detailed user profile
which records her/his background knowledge and future learning objectives and, consequently, plays
a key role in the definition of learning programs.

\item {\em It is dynamic and flexible} since it is provided with mechanisms for reacting to
variations of both user and organization exigencies and objectives.

\item {\em It is device adaptive} since it decides the typology (in particular, the ``multi-media
degree'') of learning objects to present to the user on the basis of the device she/he is currently
exploiting.

\item {\em It is generic}, i.e., it is capable of operating on a large variety of learning
contexts.

\item {\em It is XML-based}, since {\em (i)} the agent ontologies are stored as XML documents; {\em
(ii)} the communication language exploited by the various agents is ACML \cite{GrLa99}, a language
obtained by combining XML and KQML \cite{Finin*94}; {\em (iii)} the extraction of information from
the various data structures is carried out by means of {\em XQuery} \cite{XQuery}; {\em (iv)} the
manipulation of agent ontologies is performed by means of the Document Object Model (DOM)
\cite{DOM}; {\em (v)} information relative to the learning activities is represented and handled by
means of the IMS standard \cite{IMS} (see above).

\end{itemize}

\noindent This paper is organized as follows: the next section presents some preliminary
definitions; a detailed description of all agents involved in {\em X-Learn} is provided in Section
\ref{Involved-Agents}. Section \ref{Experiments} is devoted to describe a series of experiments we
have performed for testing our system performances. In Section \ref{Related-Literature} we present
related literature and compare {\em X-Learn} with various other systems already proposed in the
past. Finally, in Section \ref{Conclusions}, we draw our conclusions.

\section{Preliminaries}
\label{Preliminaries}

In this section we provide some preliminary definitions necessary to understand both the
architecture and the behaviour of {\em X-Learn}.

\begin{definition}
\label{skill} {\em A {\em skill} indicates an ability that a user wants to achieve. Examples of
skills are ``C++ programmer'', ``Webmaster'', etc. Each skill requires the knowledge of a set of
subjects. We say that a user acquires a skill when she/he knows all the subjects associated with
it. }
\end{definition}

\begin{definition}
\label{subject} {\em A {\em subject} denotes a high level topic of a skill. Examples of subjects
are ``C++ functions'', ``C++ Classes'', ``C++ Class Inheritance'', etc. Each subject may have one
or more pre-requisites; these are other subjects whose knowledge is necessary for studying it. As
an example, in order to study the subject ``C++ Class Inheritance'', it is necessary to know the
subject ``C++ Classes''. Analogously, a subject can be a pre-requisite for one or more subjects. We
say that a subject is {\em basic} if it has no pre-requisites. }
\end{definition}

\begin{definition}
\label{learning-object} {\em A {\em learning object} is an elementary learning unit relative to a
specific subject. In this paper we assume that:

\begin{itemize}

\item each learning object is relative to only one subject;

\item various learning objects could be associated with the same subject; they could differ for the
associated learning methodology, for their multimedia degree, and so on. However, all learning
objects associated with the same subject are considered equivalent from a didactic point of view.

\end{itemize}

\noindent A learning object consists of two components, namely the {\em learning object descriptor}
and the {\em learning object content}. The former describes the characteristics of the learning
object (e.g., the associated subject, the multimedia format, etc.). The latter corresponds to the
actual information content associated with the learning object and that the user must study for
learning the subject associated with it.

}
\end{definition}

As previously mentioned, subjects can be characterized by some pre-requisites which are, in their
turn, other subjects. As a consequence, a user can study a subject only if she/he knows all the
corresponding pre-requisites. We have seen that studying a subject corresponds to study one of the
learning objects associated with it. As a consequence, it is possible to introduce the concept of
learning program which allows to formally define the (partially ordered) set of learning objects
that a user must study for learning a subject starting from her/his background knowledge.

\begin{definition}
\label{learning-program} {\em A {\em learning program} $LP$ is a set of pairs of learning objects
$(LObj_s,$ $LObj_t)$ such that $LObj_t$ can be studied only after $LObj_s$ or, in other words, such
that the subject associated with $LObj_s$ is a pre-requisite of the subject relative to $LObj_t$.

Note that, in $LP$, more tuples $(LObj_{s_1}, LObj_t)$, $(LObj_{s_2}, LObj_t)$, $\ldots$,
$(LObj_{s_n},$ $LObj_t)$ could exist having $LObj_t$ as their second component; this indicates that
$LObj_t$ can be studied only after $LObj_{s_1}, LObj_{s_2}, \ldots, LObj_{s_n}$ have been learned.
In this way, $LP$ specifies also a partial order according to which the learning objects must be
studied.

}
\end{definition}

\section{The X-Learn Architecture}
\label{Involved-Agents}

\subsection{General Overview}
\label{General-Overview}

{\em X-Learn} consists of three agent typologies, namely:

\begin{itemize}

\item {\em a User-Device Agent} (hereafter $UDA$), that handles an e-learning session carried out
by a user $U$ by means of a device $D$;

\item {\em a Skill Manager Agent}, (hereafter $SMA$), that supports a user to determine the skills
of her/his interest, as well as the subjects she/he has to study for attaining a given skill, on
the basis of her/his background knowledge;

\item {\em a Learning Program Agent}, (hereafter $LPA$), that generates personalized learning
programs for a specific user $U$ needing to study a particular subject $S$, having a certain
background knowledge and exploiting a device $D$ for her/his e-learning activity.

\end{itemize}

\noindent In addition, {\em X-Learn} is provided with a Learning Object Repository ($LOR$), storing
all learning objects it handles.

As previously pointed out, the role of XML in {\em X-Learn} is crucial. Indeed:

\begin{itemize}

\item The agent ontologies are stored as XML documents; as a consequence, they are light,
versatile, easy to be exchanged and can reside on different devices and software platforms. In
spite of this simplicity, the information representation rules embodied in XML are powerful enough
to allow a sophisticated information management.

\item The agent communication language is ACML \cite{GrLa99}; this is the XML encoding of FIPA
Agent Communication Language \cite{FIPA02}. The exploitation of ACML guarantees various benefits to
{\em X-Learn}; two of the most relevant ones are the following:

\begin{itemize}

\item Developing and managing tools capable of carrying out ACML message parsing is extremely
simple; indeed, these tools can be constructed by exploiting the numerous off-the-shelf XML parsers
available over the Internet. Vice versa, in order to construct parsers for not XML-based ACL
versions, it is generally necessary to exploit a Lisp-like encoding (see \cite{GrLa99} for all
details) whose supports are more difficult to be found over the Internet.

\item Integrating Agents with a large variety of Web technologies (such as Secure Socket Layer -
SSL, for handling both the authentication of agents' identities and the encryption of ACL messages)
is very simple to be realized. Vice versa, addressing the same issues with a not XML-based Agent
Communication Language would imply heavy constraints on the agent infrastructure (think, for
example, to the great overload to be put in the ACL messages for handling these issues).

\end{itemize}

\item The extraction of information from the various data structures is carried out by means of
XQuery \cite{XQuery}. This is becoming the standard query language for the XML environment. Since
it is based on the XML framework, XQuery can handle a large data variety. It has capabilities
typical of database query languages as well as features typical of document management systems.
Finally, it is provided with various high level constructs for simplifying querying over the Web;
among them, we cite {\em constructors}, that allow the creation of XML structures within a query,
and {\em FLWR expressions}, that support iteration and variable binding.

\item The manipulation of agent ontologies is performed by means of the Document Object Model (DOM)
\cite{DOM}. This is a platform- and language-neutral interface that allows programs and scripts to
dynamically access and update the content, structure and style of XML documents. DOM makes it
possible for programmers to write applications working properly on all browsers and servers as well
as on a large variety of hardware and software platforms.

\item Learning Object Metadata are represented and handled by means of the IMS standard \cite{IMS}.
As pointed out in the Introduction, such a standard describes learning objects by means of XML
documents, validated with respect to an XML Schema. The exploitation of XML allows to manipulate
and manage learning object descriptors using the most recent XML technologies such as DOM, for data
manipulation, SAX, for data parsing, XQuery, for data querying, and so on.

\end{itemize}

\noindent In the following subsections we provide a detailed description of the various agent
typologies which {\em X-Learn} consists of.

\subsection{The User-Device Agent}
\label{UDA}

A User-Device Agent $UDA_{ij}$ is associated with a user $U_j$ exploiting a device $D_i$; it
supports $U_j$ during her/his learning activities carried out by means of $D_i$.

\subsubsection{Ontology}
\label{UDA-Ontology}

The ontology of $UDA_{ij}$ consists of a pair $\langle DP_i, UP_j \rangle$, where:

\begin{itemize}

\item $DP_i$, the Device Profile of $D_i$, stores some characteristics of $D_i$ such as the maximum
bandwidth and the medium typology (e.g., video, audio, etc.) it can handle;

\item $UP_j$, the User Profile of $U_j$, stores some characteristics of $U_j$ such as the skill
she/he wants to acquire, her/his background knowledge and the maximum time she/he can spend for a
learning program.

\end{itemize}

\noindent Table \ref{UDA-Ont-Table} illustrates the parameters characterizing the ontology of
$UDA_{ij}$ in more detail. The corresponding XML Schema is shown in Figure \ref{UDA-Ont-Schema}.

\begin{table} [t]
\begin{center}
{\scriptsize
\begin{tabular}{||l||l|l||}
\hline \hline
$DP_i$ & & the {\em Profile of the Device} $D_i$ \\
& $DId_i$ & the {\em Identifier} of $D_i$ \\
& $BMax_i$ & the {\em Maximum Bandwidth} that $D_i$ can guarantee for  \\
& & accessing or downloading data from the network \\
& $VE_i$ & the {\em Video Enabled} field of $D_i$. It is set to 1 if $D_i$ supports \\
& & video data format, 0 otherwise \\
& $AE_i$ & the {\em Audio Enabled} field of $D_i$. It is analogous to $VE_i$ but for audio \\
& $TE_i$ & the {\em Text Enabled} field of $D_i$. It is analogous to $VE_i$ but for text \\
\hline \hline
$UP_j$ & & the {\em Profile of the User} $U_j$ \\
& $UId_j$ & the {\em Identifier} of $U_j$ \\
& $DesSkill_j$ & the {\em Desired Skill} of $U_j$ (i.e., the skill $U_j$ wants to acquire) \\
& $AcqSkillSet_j$ & the {\em Acquired Skill Set} of $U_j$ \\
& $KnownSubjSet_j$ & the {\em Known Subject Set} of $U_j$. A subject $KnownSubj_{j_l}$ of \\
& & $KnownSubjSet_j$ has an identifier $SubjId_{j_l}$ and a name $SubjName_{j_l}$ \\
& $MaxTime_j$ & the {\em Maximum Time} $U_j$ can spend for a learning program \\
\hline \hline
\end{tabular}
}
\end{center}
\caption{The Ontology of $UDA_{ij}$} \label{UDA-Ont-Table}
\end{table}

\begin{figure*}[t]
{\tiny
\begin{minipage} [t] {6cm}
\begin{verbatim}

<?xml version="1.0" encoding="UTF-8"?>
<xs:schema xmlns:xs="http://www.w3.org/2001/XMLSchema">
    <!-- Definition of attributes -->
    <xs:attribute name="SubjId" type="xs:ID"/>
    <xs:attribute name="SubjName" type="xs:string"/>
    <xs:attribute name="DId" type="xs:ID"/>
    <xs:attribute name="BMax" type="xs:float"/>
    <xs:attribute name="VE" type="xs:integer"/>
    <xs:attribute name="AE" type="xs:integer"/>
    <xs:attribute name="TE" type="xs:integer"/>
    <xs:attribute name="UId" type="xs:ID"/>
    <xs:attribute name="DesSkill" type="xs:string"/>
    <xs:attribute name="MaxTime" type="xs:float"/>
    <!-- Definition of simple elements -->
    <xs:element name="AcqSkill" type="xs:string"/>
    <!-- Definition of complex elements -->
    <xs:element name="Subj">
        <xs:complexType>
            <xs:attribute ref="SubjId" use="required"/>
            <xs:attribute ref="SubjName" use="required"/>
        </xs:complexType>
    </xs:element>
    <xs:element name="AcqSkillSet">
        <xs:complexType>
           <xs:element ref="AcqSkill" minOccurs="0"
                                maxOccurs="unbounded"/>
        </xs:complexType>
    </xs:element>
    <xs:element name="KnownSubjSet">
        <xs:complexType>
           <xs:element ref="KnownSubj" minOccurs="0"
                                maxOccurs="unbounded"/>
        </xs:complexType>
    </xs:element>

\end{verbatim}
\end{minipage}
\begin{minipage}[t]{6cm}
\begin{verbatim}

    <xs:element name="DP">
        <xs:complexType>
            <xs:attribute ref="DId" use="required"/>
            <xs:attribute ref="BMax" use="required"/>
            <xs:attribute ref="VE" use="required"/>
            <xs:attribute ref="AE" use="required"/>
            <xs:attribute ref="TE" use="required"/>
        </xs:complexType>
    </xs:element>
    <xs:element name="UP">
        <xs:complexType>
            <xs:sequence>
                <xs:element ref="AcqSkillSet"/>
                <xs:element ref="KnownSubjSet"/>
            </xs:sequence>
            <xs:attribute ref="UId" use="required"/>
            <xs:attribute ref="DesSkill" />
            <xs:attribute ref="MaxTime" use="required"/>
        </xs:complexType>
    </xs:element>
    <xs:element name="UDAOntology">
        <xs:complexType>
            <xs:sequence>
                <xs:element ref="DP"/>
                <xs:element ref="UP"/>
            </xs:sequence>
        </xs:complexType>
    </xs:element>
</xs:schema>

\end{verbatim}
\end{minipage}
} \caption{The XML Schema of $UDA$} \label{UDA-Ont-Schema}
\end{figure*}

\subsubsection{Behaviour}
\label{UDA-Behaviour}

$UDA_{ij}$ is activated by $U_j$ when she/he wants to acquire a new skill. In this case $UDA_{ij}$
contacts $SMA$ and sends it the set of skills already acquired by $U_j$. In its turn, $SMA$ sends
$UDA_{ij}$ the list of skills $U_j$ might acquire; these are shown to $U_j$ who can select one of
them. When this happens, $UDA_{ij}$ adds the selected skill to $UP_j$ and the learning session
starts. In order to illustrate the exploitation of ACML, in Figure \ref{ACML-XML} we show the ACML
message that $UDA_{ij}$ sends to $SMA$. In the following, due to space limitations, we cannot
present the other ACML messages exchanged by the various agents; however, they are analogous to
that shown in Figure \ref{ACML-XML}.

\begin{figure*}[h]
{\scriptsize
\begin{minipage} [t] {6cm}
\begin{verbatim}

<?xml version="1.0" encoding="UTF-8"?>
<!DOCTYPE fipa_acl SYSTEM "fipa_acl.dtd">
<message>
    <messagetype>
        request
    </messagetype>
    <messageparameter>
        <sender link="http://www.ing.unirc.it/user">
            UDA
        </sender>
    </messageparameter>
    <messageparameter>
        <receiver link="http://www.mat.unical.it/X-learn">
            SMA
        </receiver>
    </messageparameter>
    <messageparameter>
        <ontology link="http://www.ing.unirc.it/user/UDAontology.xml">
            Ontology of UDA
        </ontology>
    </messageparameter>
    <messageparameter>
        <content>
            Request of available skills
        </content>
    </messageparameter>
    <messageparameter>
        <reply-with>
            List of skills
        </reply-with>
    </messageparameter>
</message>
\end{verbatim}
\end{minipage}
} \caption{The ACML message that $UDA$ sends to $SMA$} \label{ACML-XML}
\end{figure*}

$UDA_{ij}$ can be activated by $U_j$ also when she/he wants to continue a previously interrupted
learning program. In this case $UDA_{ij}$ exploits information stored in its ontology for
re-starting the learning program.

A learning session is carried out as follows. $UDA_{ij}$ sends to $SMA$ both the set of subjects
already known by $U_j$ and the skill she/he desires to acquire. $SMA$ identifies the subjects $U_j$
must attain for acquiring the desired skill and returns an ordered list of them to $UDA_{ij}$. The
list order reflects the pre-requisite relationships existing among subjects. At this point, $U_j$
can choose the next subject to learn.

After this, $UDA_{ij}$ contacts $LPA$ and sends it the device profile $DP_i$, the user profile
$UP_j$ and the subject $Subj_k$ that $U_j$ desires to learn. $LPA$ determines the Best Learning
Program $BLP_{ijk}$ allowing $U_j$ to learn $Subj_k$ by means of $D_i$ and sends it to $UDA_{ij}$
(see Section \ref{LPA}). This extracts each learning object of $BLP_{ijk}$ from the Learning Object
Repository and presents it to $U_j$. When she/he ends to study a learning object of $BLP_{ijk}$,
$UDA_{ij}$ updates $UP_j$ by adding the corresponding subject to the set of subjects already known
by $U_j$.

After $U_j$ has studied all learning objects of $BLP_{ijk}$, and, consequently, has acquired
$Subj_k$, she/he can decide to interrupt the learning session or, alternatively, to continue it by
studying a further subject relative to the current Desired Skill. In the former case, $UDA_{ij}$ is
de-activated; in the latter case, it contacts $LPA$ for determining the new learning program.

Finally, when $U_j$ knows all subjects associated with the current Desired Skill, $UDA_{ij}$
updates $UP_j$ by adding it to the set of acquired skills.

\subsection{The Skill Manager Agent}
\label{SMA}

A Skill Manager Agent $SMA$ supports User-Device Agents in the selection of skills and subjects to
be learned by the corresponding users.

\subsubsection{Ontology}
\label{SMA-Ontology}

The ontology of $SMA$ consists of a set of skills $SkillSet=\{Sk_1, \ldots$, $Sk_q\}$. Each skill
$Sk_l$ is characterized by a name $SkName_l$ and the list $SkSubjList_l$ of subjects to be learned
for attaining it. Subjects in $SkSubjList_l$ are ordered on the basis of the pre-requisite
relationships existing among them. The XML Schema associated with this ontology is analogous to
that relative to the ontology of $UDA_{ij}$; due to space limitations we do not show it.

\subsubsection{Behaviour}
\label{SMA-Behaviour}

$SMA$ is activated by a User-Device Agent $UDA_{ij}$ when $U_j$ wants to choose a new skill to
acquire or when she/he wants to learn a new subject relative to her/his current Desired Skill.

In the former case, $SMA$ receives from $UDA_{ij}$ the set of skills attained by $U_j$ in the past
and returns to $UDA_{ij}$ the skills present in {\em X-Learn} not yet acquired by $U_j$. The query
for skill extraction, expressed in XQuery, is shown in Figure \ref{XQuery1}.

In the latter case, $SMA$ receives from $UDA_{ij}$ the set of subjects $U_j$ already knows and the
skill she/he desires to acquire; it selects from its ontology the list of subjects necessary to be
learned for attaining the current desired skill of $U_j$, filters out those already known by $U_j$
and returns the remaining ones to $UDA_{ij}$. The associated query is illustrated in Figure
\ref{XQuery2}.

\begin{figure*}
{\scriptsize
\begin{minipage} [t] {6cm}
\begin{verbatim}
<SkillSet>
    for $i in document("http://www.mat.unical.it/X-learn/SMAOntology.xml)/*/Skill
    where empty (document("http://www.ing.unirc.it/user/UDAOntology.xml)
        /*/AcqSkillSet [AcqSkill eq $i/@Name]
        return
            <Skill>
                $i/@Name
            </Skill>
</SkillSet>
\end{verbatim}
\end{minipage}
} \caption{The query $SMA$ executes for selecting the skills present in {\em X-Learn} and not yet
acquired by $U_j$} \label{XQuery1}
\end{figure*}

\begin{figure*}[h]
{\scriptsize
\begin{minipage} [t] {6cm}
\begin{verbatim}
<SubjectSet>
    let $uda:=document("http://www.ing.unirc.it/user/UDAOntology.xml)
     let $skill:=document("http://www.mat.unical.it/X-learn/SMAOntology.xml")
        /*/Skill[Name eq $uda/*/@DesSkill]


    for $subject in $skill/SkSubjList/Subject
    where empty ($uda/*/KnownSubjSet [Subj/@Name eq $subject]
        return
            <Subject>
                $subject
            </Subject>
</SubjectSet>
\end{verbatim}
\end{minipage}
} \caption{The query $SMA$ executes for selecting the list of subjects of the current desired skill
not already known by $U_j$} \label{XQuery2}
\end{figure*}

\subsection{The Learning Program Agent}
\label{LPA}

The Learning Program Agent $LPA$ is activated by a User-Device Agent $UDA_{ij}$ whenever $U_j$
wants to study a new subject $Subj_k$. It is in charge of providing $U_j$ with a personalized
learning program for studying $Subj_k$ on the basis of her/his background knowledge and the
characteristics of the device $D_i$ she/he is currently exploiting.

\subsubsection{Ontology}
\label{LPA-Ontology}

The ontology of $LPA$ consists of a pair $\langle SubjSet, LObjSet \rangle$, where:

\begin{itemize}

\item $SubjSet$ represents the set of subjects currently available in {\em X-Learn}. Each subject
is characterized by a code, a name, the set of its pre-requisites and the set of learning objects
associated with it.

\item $LObjSet$ is the set of learning objects currently present in {\em X-Learn}. Each learning
object is characterized by an identifier, a name, the subject it refers to\footnote{Recall that a
learning object is related to only one subject but one subject might have various learning objects
associated with it.}, the URI where it can be accessed, its data format, size and duration.
Metadata for describing learning objects have been defined according to IMS specifications
\cite{IMS}. Table \ref{LPA-Ont-Table} illustrates the parameters characterizing the ontology of
$LPA$ in more detail. The corresponding XML Schema is analogous to that relative to the ontologies
of $UDA_{ij}$ and $SMA$; due to space limitations we do not show it.

\end{itemize}

\begin{table} [t]
\begin{center}
{\scriptsize
\begin{tabular}{||l||l|l||}
\hline \hline $SubjSet$ & & the {\em Set of Subjects} currently available in {\em X-Learn} \\
\hline
$Subj_l \in SubjSet$ & $SubjId_l$ & the {\em Identifier} of $Subj_l$ \\
& $SubjName_l$ & the {\em Name} of $Subj_l$  \\
& $SubjPrereqSet_l$ & the {\em Set of Pre-requisites} of $Subj_l$ \\
& $SubjLObjSet_l$ & the {\em Set of learning objects} relative to $Subj_l$ \\
\hline \hline
$LObjSet$ & & the {\em Set of learning objects} currently available at the \\
& & e-learning system \\ \hline
$LObj_m \in LObjSet$ & $LObjId_m$ & the {\em Identifier} of $LObj_m$ \\
& $LObjName_m$ & the {\em Name} of $LObj_m$ \\
& $LObjSubject_m$ & the {\em Subject} which $LObj_m$ refers to \\
& $LObjLocation_m$ & the {\em URI} where $LObj_m$ can be accessed \\
& $LObjVC_m$  & the {\em Video Component} field of $LObj_m$. It is set to 1 if \\
& & $LObj_m$ has a video component, 0 otherwise \\
& $LObjAC_m$ & the {\em Audio Component} field of $LObj_m$. It is analogous to \\
& & $LObjVC_m$ but for audio \\
& $LObjTC_m$ & the {\em Text Component} field of $LObj_m$. It is analogous to \\
& & $LObjVC_m$ but for text \\
& $LObjSize_m$ & the {\em Size}, in bytes, of $LObj_m$ \\
& $LObjDuration_m$ & the {\em Duration} of $LObj_m$. It is defined as the time, \\
& & in seconds, that $LObj_m$ takes when it is played \\

 \hline \hline
\end{tabular}
}
\end{center}
\caption{The Ontology of $LPA$} \label{LPA-Ont-Table}
\end{table}

\subsubsection{Behaviour}
\label{LPA-Behaviour}

$LPA$ is activated by $UDA_{ij}$ whenever a user $U_j$ wants to study a subject $Subj_k$ by means
of a device $D_i$. $LPA$ receives $Subj_k$, $UP_j$ and $DP_i$ from $UDA_{ij}$. It returns to
$UDA_{ij}$ the Best Learning Program $BLP_{ijk}$ allowing $U_j$ to study $Subj_k$ by means of
$D_i$. The construction of $BLP_{ijk}$ consists mainly of three steps.

\paragraph{Step 1}
During the first step $LPA$ constructs a support graph, named {\em Subject Dependency Graph}
$SDG_{jk} = \langle NS_{jk}, AS_{jk} \rangle$. $SDG_{jk}$ is constructed for guiding $U_j$ to learn
$Subj_k$ starting from basic and/or already known subjects.

As a consequence, {\em for each} list of subjects $\{Subj_1, Subj_2, \ldots, Subj_n \}$, such that:

\begin{itemize}

\item $Subj_l$ is a pre-requisite of $Subj_{l+1}$, $1 \leq l \leq n-1$;

\item $Subj_1, \ldots, Subj_n$ are not known by $U_j$;

\item $Subj_1$ is either a basic subject or a subject whose pre-requisites are already known by
$U_j$;

\item $Subj_n = Subj_k$.

\end{itemize}

\noindent $Subj_1, \ldots, Subj_n$ are added to $NS_{jk}$\footnote{In the following we shall use
the same name for indicating both a subject and the associated node in $SDG_{jk}$, when this is not
confusing.} and arcs $(Subj_1, Subj_2)$, $\ldots$, $(Subj_{n-1},$ $Subj_n)$ are added to $AS_{jk}$,
if not already present.

\paragraph{Step 2}

During the second step $LPA$ exploits $SDG_{jk}$ for determining the Best Learning Program
$BLP_{ijk}$. Such a task is carried out by suitably selecting a learning object for each subject in
$SDG_{jk}$. The learning object selection is performed according to the following guidelines:

\begin{itemize}

\item $U_j$ should exploit as much available bandwidth as possible. The available bandwidth for
$U_j$ is determined by computing the minimum between the bandwidth $BMax_i$ guaranteed by $D_i$ and
the bandwidth $BNet_j$ available on the network for $U_j$. The bandwidth required by each learning
object is computed as the ratio between its size and its duration.

\item The time required to $U_j$ to learn $BLP_{ijk}$ must be lower than $MaxTime_j$, i.e., the
maximum time $U_j$ can spend for a learning program.

\item The format of each selected learning object must be compatible with the characteristics of
$D_i$.

\item In $BLP_{ijk}$ exactly one learning object must be selected for each subject of $SDG_{jk}$.

\end{itemize}

\noindent The construction of $BLP_{ijk}$ can be properly formulated as the following optimization
problem:

\[
\begin{array}{ll}

maximize & \sum_{r = 1}^{|NS_{jk}|} \sum_{s=1}^{|SubjLObjSet_r|} \frac{LObjSize_{r_s}}{LObjDuration_{r_s}} x_{r_s} \\

s.t. & \frac{LObjSize_{r_s}}{LObjDuration_{r_s}} x_{r_s} \leq min \{BMax_i, B_{Net_j} \} \\

& \sum_{r = 1}^{|NS_{jk}|} \sum_{s=1}^{|SubjLObjSet_r|} LObjDuration_{r_s} x_{r_s} \leq MaxTime_j \\

& \sum_{s=1}^{|SubjLObjSet_r|} LObjVC_{r_s} x_{r_s} \leq VE_i , \hspace*{0.9cm} 1 \leq r \leq |NS_{jk}| \\

& \sum_{s=1}^{|SubjLObjSet_r|} LObjAC_{r_s} x_{r_s} \leq AE_i , \hspace*{0.95cm} 1 \leq r \leq |NS_{jk}|\\

& \sum_{s=1}^{|SubjLObjSet_r|} LObjTC_{r_s} x_{r_s} \leq TE_i , \hspace*{0.95cm} 1 \leq r \leq |NS_{jk}|\\

& \sum_{s=1}^{|SubjLObjSet_r|} x_{r_s} = 1 , \hspace*{3cm} 1 \leq r \leq |NS_{jk}|\\

& x_{r_s} \in \{0,1\} \\

\end{array}
\]

\noindent Here, the variable $x_{r_s}$ represents the learning object $LObj_s$ associated with the
subject $Subj_r$. $x_{r_s}=1$ if $LObj_s$ belongs to $BLP_{ijk}$.

\paragraph{Step 3}

During the third step $LPA$ simply returns $BLP_{ijk}$ to $UDA_{ij}$.

\section{Experiments}
\label{Experiments}

We have carried out various experiments for verifying the performances of {\em X-Learn}. Most of
these experiments have been conceived for verifying the capability of our system to adapt its
behaviour to both bandwidth availabilities and the characteristics of the devices exploited by
users.

In these experiments, $72\%$ of learning objects {\em available} at the Learning Object Repository
of {\em X-Learn} had a text component, $72\%$ of them had an audio component and, finally, $72\%$
of them had a video component\footnote{Remember that a learning object might contemporarily have a
text, an audio and a video component.}.

A first experiment has been performed for measuring the fraction of {\em selected} learning objects
having a text (resp., an audio, a video) component. Before carrying out the experiment we thought
that, if the available bandwidth increases, the fraction of selected learning objects having an
audio and/or a video component increases as well, whereas the percentage of selected learning
objects having a text component should be quite constant and high.

\begin{figure}[t]
\centerline{\psfig{figure=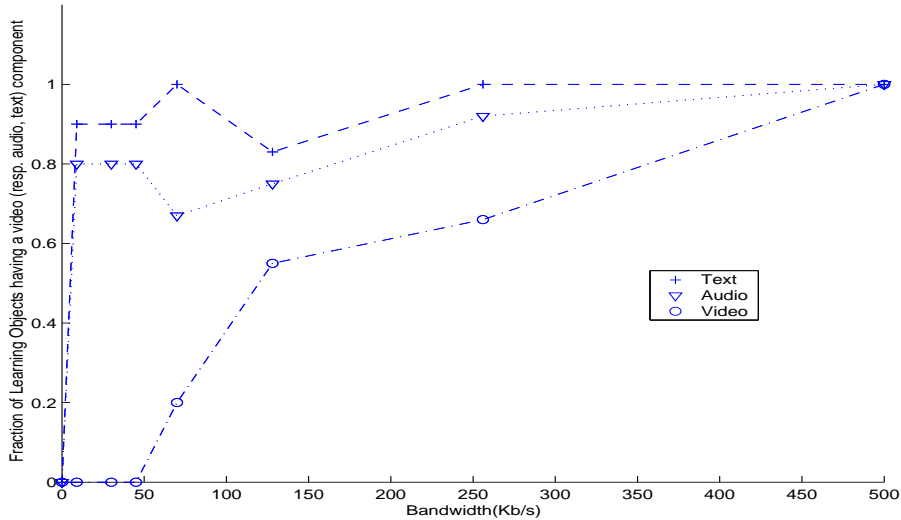,width=12cm,height=7cm}} \caption{Variation of the fraction of
selected learning objects having a video (resp., audio, text) component against the variation of available
bandwidth} \label{Esperimenti}
\end{figure}

The results we have obtained for this experiment are shown in Figure \ref{Esperimenti}. They
confirm our intuition. Indeed, it is possible to observe that:

\begin{itemize}

\item The fraction of {\em selected} learning objects having a text component is quite constant and
high; indeed, it is always greater than 80\%.

\item The percentage of {\em selected} learning objects having an audio component slightly
increases when the bandwidth increases; it is quite high, since it is always greater than 60\%.

\item In presence of a bandwidth increase, the increase of the fraction of {\em selected} learning
objects having a video component is enormous and rapid.

\end{itemize}

\noindent A second experiment has been carried out for verifying how the selection of learning
objects depends on the device exploited by the user. In this experiment, the set of available
learning objects is the same as that taken into account in the previous one.

We have considered four device typologies handling {\em (i)} text and audio, {\em (ii)} text and
video, {\em (iii)} audio and video, {\em (iv)} text, audio and video. In addition, we have
considered three situations for bandwidth availability, namely {\em (a)} low bandwidth (i.e., 9-10
kbytes/s), {\em (b)} medium bandwidth (i.e., 50-60 kbytes/s), {\em (c)} high bandwidth (i.e., over
120 kbytes/s).

Results obtained when the available bandwidth is low are shown in Table \ref{Esp-Tab1}. In this
table there is a row for each device typology; columns are associated with text, audio and video.
The element corresponding to the row ``Text and Audio'' and to the column ``Audio'' specifies the
fraction of learning objects, having an audio component, which are selected if a device handling
only text and audio is exploited. Observe that, in case of a low bandwidth, if the device can
handle text and audio (resp., video), text is preferred to audio (resp., video). Analogously, if
the device can handle video and audio, audio is preferred to video. Finally, if the device can
handle text, audio and video, video is totally filtered out, audio is partially considered whereas
text is generally selected. These results are reasonable if we consider that, in this experiment,
available bandwidth is low and video components generally require a high bandwidth.

\begin{table}
\begin{center}
{\footnotesize
\begin{tabular}{||l||c|c|c||}
\hline \hline
 & {\em Text} & {\em Audio} & {\em Video} \\
\hline \hline
{\em Text and Audio} & 0.85 & 0.60 & 0.00 \\
{\em Text and Video} & 1.00 & 0.00 & 0.00 \\
{\em Audio and Video} & 0.00 & 1.00 & 0.00 \\
{\em Text, Audio and Video} & 0.85 & 0.60 & 0.00 \\
\hline \hline
\end{tabular}
}
\end{center}
\caption{Results returned when the bandwidth is low} \label{Esp-Tab1}
\end{table}

Results returned when the available bandwidth is medium are reported in Table \ref{Esp-Tab2}.
Observe that, since available bandwidth is higher w.r.t. the previous case, the fraction of
selected learning objects having an audio and/or a video component is higher than that returned
previously.

\begin{table}
\begin{center}
{\footnotesize
\begin{tabular}{||l||c|c|c||}
\hline \hline
& {\em Text} & {\em Audio} & {\em Video} \\
\hline \hline
{\em Text and Audio} & 1.00 & 0.80 & 0.00 \\
{\em Text and Video} & 1.00 & 0.00 & 0.26 \\
{\em Audio and Video} & 0.00 & 0.65 & 0.18 \\
{\em Text, Audio and Video} & 1.00 & 0.60 & 0.20 \\
\hline \hline
\end{tabular}
}
\end{center}
\caption{Results returned when the bandwidth is medium} \label{Esp-Tab2}
\end{table}

Results obtained in presence of a high bandwidth are shown in Table \ref{Esp-Tab3}. In this case,
when the device is capable of handling text and audio, all selected learning objects have both a
text and an audio component. This is justified by considering that both text and audio require
quite a limited bandwidth. When the device handles both video and audio, generally, audio is
preferred to video even if a high percentage of selected learning objects have also a video
component. Finally, when the device handles text, audio and video, a large fraction of selected
learning objects has also an audio and/or a video component.

\begin{table}
\begin{center}
{\footnotesize
\begin{tabular}{||l||c|c|c||}
\hline \hline
 & {\em Text} & {\em Audio} & {\em Video} \\
\hline \hline
{\em Text and Audio} & 1.00 & 1.00 & 0.00 \\
{\em Text and Video} & 1.00 & 0.00 & 0.75 \\
{\em Audio and Video} & 0.00 & 0.95 & 0.65 \\
{\em Text, Audio and Video} & 1.00 & 0.90 & 0.65 \\
\hline \hline
\end{tabular}
}
\end{center}
\caption{Results returned when the bandwidth is high} \label{Esp-Tab3}
\end{table}

\section{Related Literature}
\label{Related-Literature}

The convergence of mobile communications and handheld computers offers new interesting
opportunities in e-learning activities; in this section we focus on some adaptive e-learning
systems and we try to highlight their similarities and differences w.r.t our approach. More details
on adaptive e-learning systems can be found in \cite{Brusilovsky99}.

In \cite{ShCoWe02} the authors propose an handheld learning device and an appropriate software
infrastructure to support children education. The main components of the proposed architecture are:
{\em (i) a learning manager}, which stores a local cache of learning objects extracted by a
repository and exploits specific software agents to search and organize learning objects, {\em (ii)
a communication manager}, which creates direct voice and data communication channels for
disseminating learning materials and handles resource sharing.

Similarly to {\em X-Learn}, \cite{ShCoWe02} develops a technology for assisting individuals and
groups to {\em learn anytime} and {\em anywhere}; in addition, in both the approaches, learning
materials follow the IMS standard and might have different multimedia formats. In spite of these
similarities, the approach of \cite{ShCoWe02} and {\em X-Learn} appear complementary; indeed, in
\cite{ShCoWe02}, the authors modify an existing handheld device to support learning activities
whereas {\em X-Learn} adapts the learning objects distribution to the device characteristics.

In \cite{RaFrKa02} a multi-agent prototype called $CITS$ (Confidence Intelligent Tutoring Agent) is
proposed. $CITS$ approach aims at being {\em adaptive} (i.e., it can adjust learning materials to
meet user needs) and {\em dynamic} (i.e., it adapts the offered service to user current behaviour).
$CITS$ architecture consists of five kinds of agents, namely: {\em (i)} a {\em Cognitive Agent},
that creates a model for each learner, representing her/his level and learning style; {\em (ii)} a
{\em Behaviour Agent}, that monitors learner behaviour during her/his interaction with the system
for improving the model produced by the Cognitive Agent; {\em (iii)} a {\em Guide Agent}, that
selects and classifies information potentially useful for the learner; {\em (iv)} an {\em
Information Agent}, that searches over the Internet for extra information required by the learner
and, {\em (v)} a {\em Confidence Agent}, that is in charge of strengthening the confidence between
the learner and the system. In $CITS$ learning information is fragmented in simple pieces called
{\em knowledge targets}; these might have different multimedia formats.

Both $CITS$ and {\em X-Learn} are XML-based multi-agent systems and both of them support the
dissemination of learning materials having different multimedia formats. The main differences
existing between them are the following: {\em (i)} $CITS$ knowledge targets and {\em X-Learn}
learning objects are different in their characteristics and purposes; {\em (ii)} $CITS$ offers more
``freedom degrees'' in the learning program definition; {\em (iii)} $CITS$ does not support device
adaptivity.

\cite{BrRi02} proposes a device-aware e-learning system as a part of a more complex e-learning
platform, named {\em KnowledgeSea}. The core of the system proposed in \cite{BrRi02} is a {\em
self-organized hyperspace map}, i.e. an automatically-built map that provides a concise navigation
support for a relatively large learning hyperspace. The map may help a user to find and access
on-line educational resources by means of mobile wireless devices.

The approach of \cite{BrRi02} is quite similar to {\em X-Learn}; indeed, both of them take into
account the device a user is exploiting for accessing educational resources. The main differences
existing between them are the following: {\em (i)} the self-organized hyperspace map provides a
more flexible mechanism for selecting learning objects; however, it does not handle pre-requisite
relationships possibly existing among learning objects; {\em (ii)} \cite{BrRi02} does not handle
the construction of a complete learning program; vice versa, in {\em X-Learn}, $LPA$ has been
conceived exactly for this purpose.

In \cite{ShShCh01} the authors propose $IDEAL$ (Intelligent Distributed Environment for Active
Learning), a multi-agent system for {\em active distance learning}. $IDEAL$ consists of: {\em (i)}
a {\em personal agent}, handling the profile (i.e., the background knowledge, the interests and the
learning style) of a learner; {\em (ii)} a {\em course agent}, managing both the materials and the
teaching technique of a course; {\em (iii)} a {\em teaching agent}, behaving as an intelligent
tutor for a learner. In $IDEAL$, course materials are decomposed into small components called {\em
Lecturelets}. These are XML documents containing JAVA code; they are dynamically assembled to cover
course topics according to learner progress.

$IDEAL$ and {\em X-Learn} share various similarities; indeed, both of them are XML based and
exploit user modeling techniques. The main differences existing between them are the following:
{\em (i)} the {\em Curriculum Sequencing Activity} of $IDEAL$ and the {\em Best Learning Program}
construction of {\em X-Learn} are based on different philosophies and strategies; {\em (ii)}
$IDEAL$ exploits non-standard and complex constructs for managing course contents (i.e.
LectureLets) whereas {\em X-Learn} uses the concept of learning object, derived from IMS standard.

In \cite{Zaiane02} an approach for exploiting {\em web-mining} techniques to build a software agent
supporting e-learning activities is presented. The proposed agent acts as a recommender system,
i.e. it can produce both {\em suggestions} (helping the learner to better navigate through on-line
materials) and {\em shortcuts} (helping the learner to quickly find needed resources). In order to
perform all these activities, the system intensively exploits a user profile taking into account
learner access history. {\em X-Learn} and \cite{Zaiane02} share some important features; in
particular, both of them exploit a user profile and operate by constructing the most appropriate
learning program. The main differences existing between {\em X-Learn} and \cite{Zaiane02} are the
following: {\em (i)} \cite{Zaiane02} is a {\em single-agent} architecture whereas our approach is
{\em multi-agent}; {\em (ii)} the learning program construction is based on data mining techniques
in \cite{Zaiane02}, whereas is performed by means of graph-based strategies in {\em X-Learn}.

In \cite{SiVi02} the system $ELETROTUTOR$ is proposed; this is a multi-agent system implemented on
a JADE platform. $ELETROTUTOR$ consists of the following agents: {\em (i)} a {\em Pedagogical
Agent}, performing learning activities, such as the distribution and the dissemination of examples
and exercises; {\em (ii)} a {\em Remote Agent}, managing the communication between the learner and
the system; {\em (iii)} a {\em Communication Agent}, handling agent communications, and {\em (iv)}
a {\em Student Model Agent}, handling a student profile and exploiting it for performing the
learning activities. Both {\em X-Learn} and $ELETROTUTOR$ are {\em multi-agent systems} and both of
them adapt the dissemination of learning contents to user profiles. As for differences between
$ELETROTUTOR$ and {\em X-Learn}, we observe that the former does not handle device adaptivity and
multimedia information that are, instead, managed by the latter.

\section{Conclusions}
\label{Conclusions}

In this paper we have proposed {\em X-Learn}, an XML-based multi-agent system for supporting
e-learning activities.

We have seen that, in {\em X-Learn}, three typologies of agents are present, namely {\em (i)} a
{\em User-Device Agent}, that handles an e-learning session carried out by a user $U$ by means of a
device $D$; {\em (ii)} a {\em Skill Manager Agent}, that supports a user $U$ to determine the
skills and the subjects she/he has to study; {\em (iii)} a {\em Learning Program Agent}, that
generates personalized learning programs for a specific user $U$ needing to study a particular
subject $S$, having a certain background knowledge and exploiting a device $D$ for her/his learning
activities.

We have shown that {\em X-Learn} is adaptive w.r.t. the profile of both the customer and the device
she/he is exploiting for carrying out the learning activities. Finally, we have seen that it is
XML-based since: {\em (i)} the agent ontologies are stored as XML documents; {\em (ii)} the
communication language exploited by the various agents is ACML; {\em (iii)} the extraction of
information from the various data structures is carried out by means of XQuery; {\em (iv)} the
manipulation of agent ontologies is performed by means of DOM; {\em (v)} learning objects are
represented and handled by means of IMS standard.

As for future work, we plan to study the possibility to enrich the proposed multi-agent model with
other features capable of improving its effectiveness and completeness in supporting a large
variety of activities related to e-learning. As an example, it might be interesting to define
various learning strategies to allow a user to specify the preferred learning strategy and,
finally, to consider such a preference when the Best Learning Program is constructed.

As a second improvement, particularly interesting when {\em X-Learn} is exploited for managing
employee learning in an organization, it could be possible to define career paths for the various
employees and to relate learning programs with them.

As a final extension, it could be possible to provide {\em X-Learn} with a team building
functionality capable of assigning employees to project teams on the basis of the skills acquired
during e-learning activities.

\end{document}